\documentclass[
aps,
prl,
twocolumn,
showpacs,
superscriptaddress,
preprintnumbers,
amsmath,
amssymb,
floatfix,
10pt,
reprint]{revtex4-1}

\usepackage{graphicx}
\usepackage{dcolumn}
\usepackage{bm}
\usepackage{multirow}
\usepackage{natbib}
\usepackage{xspace}
\usepackage[colorlinks=true,allcolors=blue]{hyperref}

\usepackage[usenames,dvipsnames,svgnames,table]{xcolor}
\usepackage[normalem]{ulem}

\usepackage[caption=false]{subfig}

\usepackage[T1]{fontenc}
\usepackage{txfonts}
\usepackage[utf8]{inputenc}

\usepackage{wasysym}
\usepackage{amsmath}
\usepackage{amssymb}

\newcommand{\mk}{\textrm{mk}}
\newcommand{\bk}{\textrm{bk}}
\newcommand{\cc}{\textrm{cc}}
\newcommand{\LR}{\textrm{LR}}
\newcommand{\QS}{\textrm{QS}}
\newcommand{\MD}{\QS}
\newcommand{\gcc}{\gamma^{\textrm{\cc}}}
\newcommand{\ecc}{\epsilon^{\textrm{\cc}}}
\newcommand{\dz}{\Delta z}

\newcommand{\gammacc}{\ensuremath{\gamma^{\textrm{cc}}}\xspace}

\begin{document}


\title{Contact Changes near Jamming}

\author{Merlijn S. \surname{van Deen}}
\email[]{deen@physics.leidenuniv.nl}
\affiliation{Huygens-Kamerlingh Onnes Lab, Universiteit Leiden, Postbus
9504, 2300 RA Leiden, The Netherlands}

\author{Johannes Simon}
\affiliation{Huygens-Kamerlingh Onnes Lab, Universiteit Leiden, Postbus 9504, 2300 RA Leiden, The
Netherlands}

\author{Zorana Zeravcic}
\affiliation{School of Engineering and Applied Sciences, Harvard University, Cambridge, Massachusetts 02138, United States}

\author{Simon Dagois-Bohy}
\affiliation{Huygens-Kamerlingh Onnes Lab, Universiteit Leiden, Postbus 9504, 2300 RA Leiden, The
Netherlands}

\author{Brian P. Tighe}
\affiliation{Process \& Energy Laboratory, Delft University of Technology, Leeghwaterstraat 39, 2628 CB Delft, The Netherlands}

\author{Martin \surname{van Hecke}}
\email[]{hecke@physics.leidenuniv.nl}
\affiliation{Huygens-Kamerlingh Onnes Lab, Universiteit Leiden, Postbus 9504, 2300 RA
Leiden, The Netherlands}

\date{\today}

\begin{abstract}
We probe the onset and effect of contact changes in soft harmonic particle packings which
are sheared quasi-statically. We find that the first contact changes are
the creation or breaking of contacts on a \emph{single} particle. We characterize the critical strain, statistics of breaking versus making a contact, and ratio of shear modulus before and after such events, and
explain their finite size scaling relations. For large systems at finite pressure,
the critical strain vanishes but the ratio of shear modulus before and after a contact change
approaches one: linear response remains relevant in large systems.
For finite systems close to jamming the critical strain also vanishes, but here linear response already breaks down after a single contact change.
\end{abstract}

\maketitle

Exciting progress in capturing the essence of the jamming transition
in disordered media such as emulsions, granular matter, and foams has been made
by considering the linear response of weakly compressed packings of repulsive, soft particles. 
When the confining pressure $P$ approaches its critical value at zero, the resulting unjamming transition bears hallmarks of a critical phase transition:
properties such
as the contact number and elastic moduli exhibit power law scaling
\cite{bolton1990rigiditylosstransition, durian1995foammechanicsat,lacasse1996modelelasticitycompressed,
ohern2002randompackingsfrictionless,
ohern2003jammingatzero,wyart2005rigidityamorphoussolids,katgert2010jammingandgeometry,tighe2011relaxationsandrheology},
time and length scales diverge \cite{ohern2003jammingatzero,
wyart2005geometricoriginexcess,silbert2005vibrationsanddiverging,
ellenbroek2006criticalscalingin}, the material's response becomes singularly non-affine \cite{ellenbroek2006criticalscalingin, ellenbroek2009jammedfrictionlessdisks}
and finite size scaling governs the behavior for small numbers of particles $N$ and/or small $P$
\cite{dagoisbohy2012softspherepackings,goodrich2012finitesizescaling, goodrich2013jamminginfinite}.

However, one may question the validity of linear response for  athermal amorphous solids
\cite{combe2000strainversusstress,hentschel2011doathermalamorphous,schreck2011repulsivecontactinteractions}.
Due to disorder, one
expects local regions arbitrary close to failure, and in addition, near their critical point 
disordered solids are extremely fragile --- even a tiny perturbation may lead to an intrinsically nonlinear response \cite{wyart2012marginalstabilityconstrains,olsson2007criticalscalingshear, tighe2010modelscalingstresses,schreck2011repulsivecontactinteractions, gomez2012uniformshockwaves,gomez2012shocksnearjamming,
wildenberg2013shockwavesin}.
To avoid such subtleties, 
numerical studies of linear response have either resorted to
simulations with very small deformations (strains of $10^{-10}$ are not uncommon in such studies \cite{schreck2010experimentalandcomputational}), or have focused on the {\em strict} linear response extracted from the
Hessian matrix \cite{ellenbroek2006criticalscalingin, ellenbroek2009jammedfrictionlessdisks,dagoisbohy2012softspherepackings,
goodrich2013jamminginfinite}.

\begin{figure}[bt!]
\centering
\includegraphics{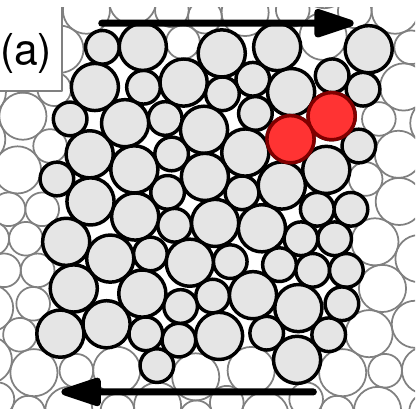}
\includegraphics{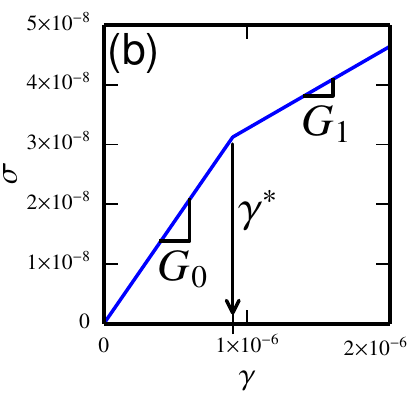}
\caption{(color online).
(a) The first contact change in a sheared packing  ($N=64$, $P=10^{-6}$) occurs
at a strain $\gamma^* = 9.003851(2)\times 10^{-7}$, when the two marked particles lose their contact. (b) The corresponding stress-strain curve remains continuous but exhibits a
sharp kink; we define $G_0$ as the shear modulus of the undeformed packing, and $G_1$ as the shear modulus of the packing just above $\gamma^*$.
\label{fig:pack-en-stresstrain}}
\end{figure}

Here we probe the first unambiguous deviations from strict linear response:  contact changes under quasistatic shear (Fig.~\ref{fig:pack-en-stresstrain}a).
We focus on three questions:
{\em (i)} What is the mean strain $\gammacc$ at which the first
contact change arises? $\gammacc$ should vanish when either $N$ diverges or $P$ vanishes.  We find a novel finite size scaling relation for $\gammacc$, where $\gammacc \sim P$ for small systems close to jamming ($N^2P \ll 1$), and $\gammacc \sim \sqrt{P}/N$ for $N^2P \gg 1$.
{\em{(ii)}} What is the nature of the first contact changes?
Plastic deformations under shear have been studied extensively in systems far from jamming, which display avalanches: collective, plastic events in which multiple contacts are broken and formed and the stresses exhibit discontinuous drops \cite{maloney2006amorphoussystemsin, lemaitre2007plasticresponsetwo, salerno2012avalanchesinstrained,hentschel2010sizeplasticevents,manning2011vibrationalmodesidentify}. A few studies have focused at what happens for hard particles, in the singular limit  
where even a single contact break may induce
a complete loss of rigidity \cite{combe2000strainversusstress,hentschel2011doathermalamorphous,wyart2012marginalstabilityconstrains}.
In contrast, we find that {\em near} jamming the first events are the making or breaking of a
single contact, and that the stress remains continuous.
The probabilities for contact making and breaking are governed by finite size scaling, with making and breaking equally likely for $N^2P \gg 1$, but contact breaking dominant for $N^2P \ll 1$.
{\em (iii)}
How do contact changes affect linear response? For finite systems close to jamming, even a single contact change can strongly affect the elastic response (Fig.~\ref{fig:pack-en-stresstrain}b). Clearly,
calculations based on the Hessian matrix of
the undeformed packing are then no longer strictly valid.
As a result, the relevance of the linear response scaling relations
are currently under dispute
for systems close to jamming, at finite temperature, or in the thermodynamic limit  \cite{schreck2011repulsivecontactinteractions, goodrich2013commentrepulsivecontact,schreck2013responsetocomment,ikeda2013dynamiccriticalityat,wang2013criticalscalingin,goodrich2014whendojammed}.
By comparing the shear modulus before ($G_0$) and after ($G_1$) the first contact change, we find that
their ratio again is governed by finite size scaling, and while the ratio
$G_1/G_0$ approaches 0.2 for small $N^2P$, for large $N^2P$,
$G_1/G_0 \rightarrow 1$.

Our work suggests that while the range of {\em strict} validity of linear response vanishes for small $P$ and large $N$, macroscopic
quantities such as the shear modulus are relatively insensitive to contact changes
as long as $P \gg 1/N^2$. Hence, linear response quantities remain relevant for finite $P$ and large $N$, while for $P \ll 1/N^2$, a single contact change already changes the
packing significantly. The qualitative differences in
the nature of contact changes close to and far from jamming suggests that plasticity, creep, and flow \emph{near} jamming are controlled by fundamentally different mechanisms than plastic flows in systems \emph{far from} jamming \cite{tighe2010modelscalingstresses,salerno2012avalanchesinstrained,hentschel2010sizeplasticevents,maloney2006amorphoussystemsin,olsson2007criticalscalingshear,manning2011vibrationalmodesidentify}.

\textit{Protocol: } We generate {\em shear stabilized} 2D packings of $N$ soft harmonic particles with unit spring constant as described in \cite{dagoisbohy2012softspherepackings}. Such shear stabilized packings
are guaranteed to have a strictly positive shear modulus $G_0$ and, moreover,
have zero residual shear stress \cite{dagoisbohy2012softspherepackings}.
As \gammacc is expected to vanish for large $N$, finite size analysis is crucial, necessitating a wide range of system sizes --- here $N$ ranges from $16$ to $4096$ and we vary $P$ from $10^{-7}$ to $10^{-2}$.

To detect contact changes, we repeatedly impose small simple shear deformations $\Delta\gamma$ at constant volume
and let the system relax. When a change in the contact network is detected between strains $\tilde\gamma$ and $\tilde\gamma + \Delta\gamma$,  we determine the precise
strain at the first contact change, $\gamma^*$, by bisection (i.e., going back to the system at
$\tilde\gamma$, dividing $\Delta\gamma$ by two, etc),
resulting in an accuracy $\Delta\gamma / \gamma^* < 10^{-6}$.

\begin{figure}[tb!]
\centering
\includegraphics{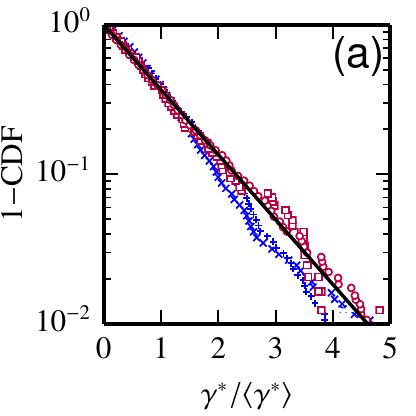}
\includegraphics{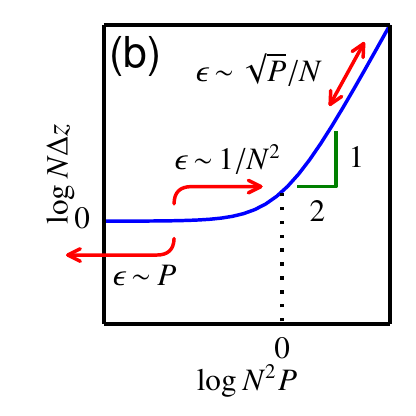}
\caption{(color online). (a) Complementary cumulative distribution function (ccdf) of $\gamma^*/\langle \gamma^* \rangle$ for $N=16; P=10^{-6}$ ($\times$), $N=16; P=10^{-2}$ ($+$), $N=1024; P=10^{-6}$ ($\square$), and $N=16; P=10^{-2}$ ($\circ$). The black line is the ccdf for an exponential distribution with unity mean.
(b) Scaling of $N\Delta z /2$ as a function of $N^2P$ (data from earlier simulations \cite{dagoisbohy2012softspherepackings}). The arrows indicate volumetric strains corresponding to a single contact change.}
\label{CDF-and-dZ-scaling}
\end{figure}

\begin{figure*}[t!b]
\centering
\includegraphics{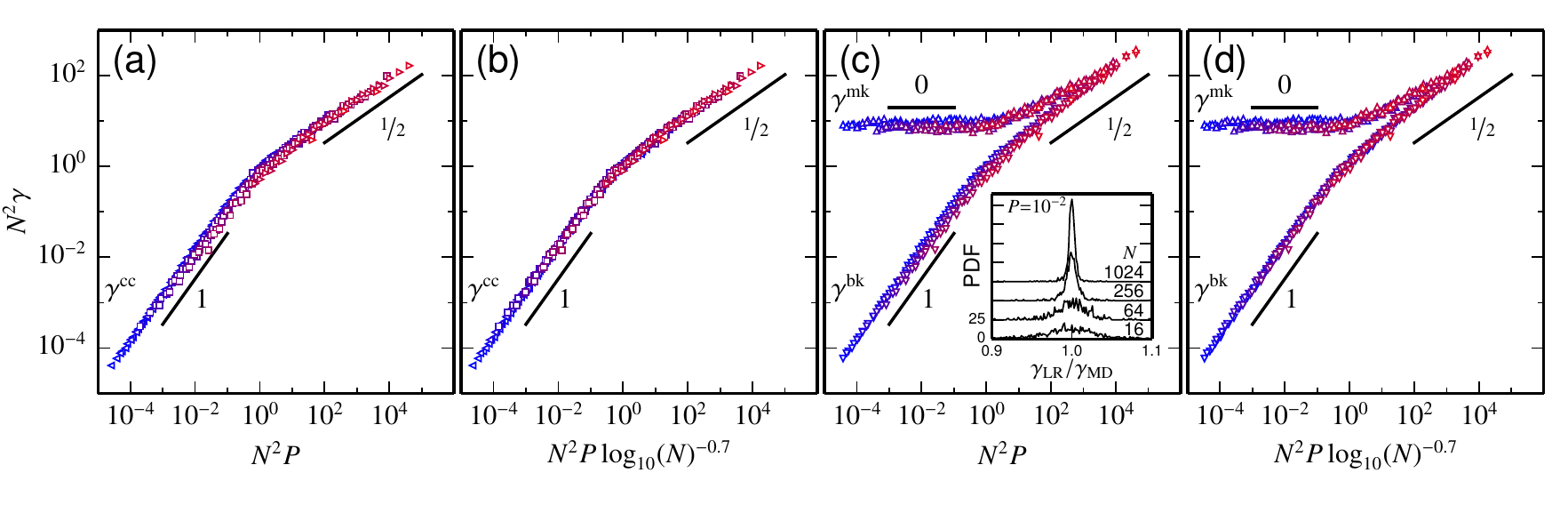}
\caption{(color online). Scaling of $\gamma^{\textrm{cc}}$, $\gamma^{\textrm{bk}}$ and $ \gamma^{\textrm{mk}}$.
(a) Scaling of the strain at first contact change.
(b) Log corrections improve the  collapse.
In (a) and (b), symbols indicate packing sizes: $\lhd\ (N \leq 32)$, $\square\ (32 < N \leq 1024)$ and $\rhd\ (N > 1024)$.
(c) Scaling of the strains for contact making ($\bigtriangleup$) and contact breaking ($\bigtriangledown$).
(c, inset) PDF of $\gamma_{\LR} / \gamma_{\MD}$ for $P=10^{-2}$ and various $N$, to compare the characteristic strains in strict linear response, $\gamma_\LR$, to those from quasistatic shear simulations, $\gamma_\MD$. Curves have been offset for clarity.
(d) Again, log corrections improve the collapse.}
\label{fig:gamma_lr_vs_gamma_md}
\label{fig:gammamed_collapse}
\end{figure*}

The first contact changes come in different flavors, and we can distinguish isolated contact making or breaking, multiple contact making or multiple breaking events, and
mixed events where contacts are both broken and created. In all these cases, rattlers
need to be treated carefully. First, in approximately 1\% of pure contact making events, a rattling particle becomes non-rattling, leading to the creation of three load bearing contacts. As these events depend on the ill-defined original location of the rattling particle, they are not included in the analysis.
Secondly, a substantial fraction of contact breaking events (10-20\%) leads to creation of rattlers, where not one but three contacts are broken simultaneously. This large proportion is not surprising, as weak contacts can easily be broken and are associated preferentially with near-rattling particles.
These events, which are well-defined, are included in our statistics.
Finally, mixed events start to play a role at high pressures, but even at $P=0.01$ less than $5\%$ of the first events are composite, and their likelihood rapidly vanishes at lower pressures; Therefore we will not include these in our analysis.
In the remainder of this Letter we focus on the statistics of the first
contact making or breaking event.

\textit{Characteristic Strain: }
We find that for fixed $P$ and $N$, the probability distribution of the
strain $\gamma^*$ at which the first contact making or breaking event arises
closely resembles an exponential distribution.
To show this, we have determined for all $P$ and $N$ the
complementary cumulative distributions (which are also exponential), and in
Fig.~\ref{CDF-and-dZ-scaling}(a) we plot four representative cases. Their exponential nature
implies that contact changes under shear can be seen as a Poisson process, and we
define $\gcc$ as the ensemble average of $\gamma^*$. We note that while the underlying rate $\sim 1/\gcc$ is constant up
to the first contact change, this rate can and will change beyond the first contact change.

As expected, we find that \gammacc increases with $P$ and decreases for larger $N$.
The question then arises: At what strain do we expect the first contact change? To start answering this, let us first consider changes in volume to derive a characteristic volumetric strain $\ecc$ for the first contact change, in both small and thermodynamically large systems. We then demonstrate numerically that the same characteristic strain governs shear.

In Fig.~\ref{CDF-and-dZ-scaling}(b) we sketch the scaling of the excess contact number $\dz$ with pressure $P$, based on data reported in Refs.~\cite{dagoisbohy2012softspherepackings,
goodrich2012finitesizescaling,
goodrich2013jamminginfinite}. The scaling relation in the thermodynamic limit is well known, $\Delta z \sim \sqrt{P}$ \cite{durian1995foammechanicsat}. It is convenient to rewrite it in the extensive form $ N \Delta z \sim \sqrt{N^2 P}$.
Making or breaking a contact increases or decreases $N\Delta z/2$ by one, and the associated change in pressure $\delta P$ can be determined from
$N \Delta z/2 \pm 1 \sim \sqrt{N^2 (P \pm \delta P)}$. The typical volumetric strain $\ecc = \delta P/K \sim \delta P$ \footnote{The bulk modulus $K$ is $O(1)$ in packings of soft harmonic spheres \cite{ohern2003jammingatzero}.} so we obtain $\ecc \sim \sqrt{P}/N $.

The small system limit is different, as $N \Delta z$  reaches a plateau independent of $P$ and $N$ -- the system is one contact away from losing rigidity \cite{dagoisbohy2012softspherepackings, goodrich2012finitesizescaling, goodrich2013jamminginfinite}, as illustrated in Fig.~\ref{CDF-and-dZ-scaling}(b).
Hence contacts can only break when $P\rightarrow 0$, and the typical strain needed to break the last contact is $\epsilon^{\bk} \sim P$.
The strain to create an additional contact, $\epsilon^{\mk}$, follows from the crossover between the two branches of $N\Delta z$ in Fig.~\ref{CDF-and-dZ-scaling}(b), so that $\epsilon^{\mk} \sim 1/N^2$. The characteristic strain for the first contact \emph{change}, $\epsilon^{\cc}$, will be dominated by the smallest of the strains $\epsilon^{\mk}$ and $\epsilon^{\bk}$. As for small systems $N^2P \ll 1$, it follows that
$\epsilon^{\bk} \ll \epsilon^{\mk}$, so that contact breaking will dominate for small systems.

In summary,  the characteristic strains under volumetric strain are predicted to be
\begin{equation}\begin{array}{rrccc}
~                  & ~ & \epsilon^{\bk} & \epsilon^{\mk} & \epsilon^{\cc}\\
\multirow{2}{*}{\ensuremath{\epsilon \sim\ \bigg\{}} & N^2 ~P \ll 1:& P           & 1/N^2       & P,\\
                   & N^2~P \gg 1:& \sqrt{P}/N  & \sqrt{P}/N  & \sqrt{P}/N.
\end{array}\label{eqn:scaling}\end{equation}
It follows that $N^2\ecc$ will collapse when plotted as function of $N^2 P$.

In Fig.~\ref{fig:gammamed_collapse}(a) we plot our rescaled data for $\gammacc$, i.e.~for {\em sheared} packings. Surprisingly,
the scalings predicted for volumetric deformations \emph{also} describe the characteristic strains for shear!
Moreover, our collapsed data exhibits the two scaling regimes predicted in Eq.~(\ref{eqn:scaling}) for large and small values of $P N^2$.

We note that the data collapse of $N^2 \gcc$ vs $N^2 P$ is good but not excellent. However,
there is mounting evidence that
the upper critical dimension of jamming is two, and several recent
accurate simulations of 2D systems near jamming show similar concomitant deviations from pure scaling \cite{tighe2010modelscalingstresses,
tighe2011relaxationsandrheology,
goodrich2013jamminginfinite,
dagoisbohy2014oscillatoryrheologynear}. As recently determined for the scaling of the contact number in 2D, such corrections take the form of log corrections to
$N^2P$ of the form $N^2P \log(N)^{-\beta}$, with $\beta \approx 0.7$ \cite{goodrich2013jamminginfinite}. Inspired by this, we replot our data for
$\gcc$ as a function of $N^2 P \log(N)^{-0.7}$, and obtain very good data collapse
(Fig.~\ref{fig:gammamed_collapse}b). We conclude that the simple scaling arguments put forward in Eq.~(\ref{eqn:scaling}) capture the scaling of $\gcc$.

\textit{Making versus Breaking: } Our scaling argument makes separate predictions for the characteristic strains of the first creating and first destruction of contacts,
but these are hard to determine independently in numerics. For example, for
small $N^2  P$ we predict that $\gamma^{\mk} \gg \gamma^{\bk}$, but that means that
almost all first contact change events are contact {\em breaking}, and even if we
observe a few contact creations (in particular when breaking events occur at atypically large strains), there is a dependency between making and breaking events
that cannot be disentangled in direct simulations.

To gain access to $\gamma^{\mk}$ and $\gamma^{\bk}$ independently, we
use the fact that contact changes can be predicted from strict linear response.
We start by extracting the linear prediction for the particle displacements $\delta x_i$ under shear from the Hessian matrix as $\Delta x_i = \gamma u_i$ \cite{ellenbroek2006criticalscalingin,ellenbroek2009jammedfrictionlessdisks,maloney2006amorphoussystemsin,manning2011vibrationalmodesidentify,dagoisbohy2012softspherepackings,tighe2011relaxationsandrheology}. We then combine this with the overlaps and gaps between particles $i$ and $j$ at $\gamma=0$,
 and determine the strain $\gamma_{ij}$ at which  contact $ij$ is predicted to break or close. The minimum of $\gamma_{ij}$ for all particle pairs in contact determines $\gamma^{\bk}$, while the minimum for all pairs not in contact determines $\gamma^{\mk}$. The minimum of both then determines $\gamma^*$.

The correspondence between
the value of $\gamma^*$ obtained from quasistatic simulations and $\gamma^*$ obtained from linear response is excellent, with an error smaller than 10\% in the worst case scenario, and typically smaller than 1\% (Fig.~\ref{fig:gamma_lr_vs_gamma_md}c, inset). In the vast majority of cases we also identify the correct contact, and whether it breaks or is created; in the case of the creation of a rattler, linear response predicts a tightly bunched triplet of $\gamma_{ij}$'s. Hence, strict linear response predicts its own demise.

The correspondence between quasistatic simulations and linear response also indicates that contact changes are the dominant source of nonlinearity (versus geometric effects).
Using linear response we can thus calculate the strains
where the first contact is created or broken and determine their mean values $\gamma^{\bk}$ and $\gamma^{\mk}$ as function of $N$ and $P$.

In Fig.~\ref{fig:gammamed_collapse}(c) we show the variation of $N^2 \gamma^{\bk}$ and $N^2 \gamma^{\mk}$ with $N^2 P$, which confirms all the predicted scalings in Eq.~(\ref{eqn:scaling}): for large $N^2 P$, $\gamma^{\mk}$ approaches $\gamma^{\bk}$ and scales as $\sqrt{P}/N$, whereas
for small $N^2 P$, $\gamma^{\mk}$  scales as $1/N^2$, whereas $\gamma^{\bk} \sim P$. As before, the data collapse is reasonably good, and gets improved by the aforementioned log-corrections (Fig.~\ref{fig:gammamed_collapse}d).

\textit{Effect of a single contact change: } What happens for strains larger that $\gamma^*$? It has been suggested that, for purely repulsive particles, linear response is no longer valid for large $N$ \cite{schreck2011repulsivecontactinteractions}, leading
to a lively debate
\cite{goodrich2013commentrepulsivecontact,schreck2013responsetocomment,goodrich2014whendojammed}.
On the one hand, it is clear that for small systems, the breaking or creation of a single contact can have a substantial effect, in particular close to jamming (see Fig.~\ref{fig:pack-en-stresstrain}b) --- but what happens for larger systems? Our data implies that $\gamma^*$ vanishes in the thermodynamic limit, so the question is what, then, is the relevance of linear response quantities?

\begin{figure}[t!]
\centering
\includegraphics{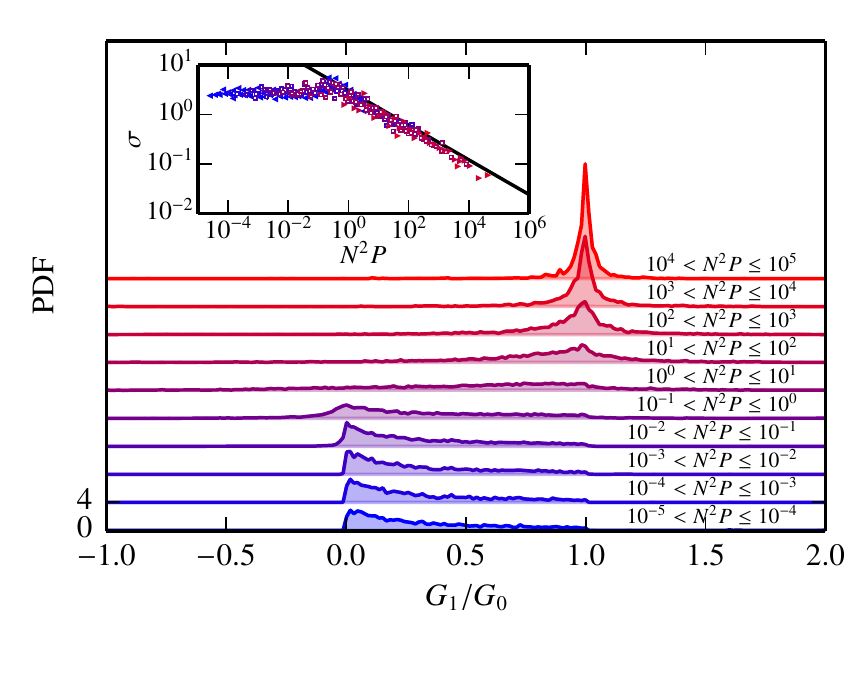}
\caption{(color online). The probability distribution functions of $G_1/G_0$ for a range of values of $N$ and $P$ become narrowly peaked when $N^2 P$ becomes large. We offset curves for different $N^2P$ for clarity. Inset: the standard deviation $\sigma$ of the distribution of $G_1/G_0$ vanishes
as $(N^2 P)^{-\beta}$, with $\beta = 0.35 \pm 0.01$, as indicated by the fitted line.
 \label{fig:sigma_and_gamma}}
\end{figure}

To probe the relevance of linear response, we determined the distribution  $P(G_1/G_0)$, where $G_0$ and $G_1$ denote the shear modulus before and after the first contact change (Fig.~\ref{fig:sigma_and_gamma}). We find that the shape of these
distributions varies widely and is determined by $N^2 P$. We can distinguish three regimes: {\em(i)} For $N^2 P \ll 1$,
$G_1  < G_0$ and $\langle G_1/G_0 \rangle \approx 0.2$. The signs of
$G_0$ and $G_1$ are both positive in this regime. $G_0$ has to be positive as we use SS packings \cite{dagoisbohy2012softspherepackings}. The sign of $G_1$ is not immediately obvious, but we note that for it to become negative, a finite prestress
is needed, but for $P\rightarrow0$ this prestress vanishes so that $G_1$ remains positive here \cite{wyart2005geometricoriginexcess,goodrich2013jamminginfinite}.
{\em(ii)} For $N^2 P \approx 1$, the
prestresses become important, but as the number of excess contacts is still small, $G_1$ now can become negative. Indeed we find that
$P(G_1/G_0)$ has a wide distribution which now acquires a finite weight for negative $G_1/G_0$.
{\em(iii)} For $N^2 P \gg 1$, $G_1$ approaches $G_0$, and the distribution $P(G_1/G_0)$ becomes sharper with increasing $N^2P$.
This can be understood by noting that
for $N^2 P \gg 1$, making and breaking of contacts is equally likely, and that $G$ varies as $\Delta z$. As the width of $P(G_1/G_0)$ scales as the difference in $G_1/G_0$ when either a contact is added or removed, we
estimate the values of $G_1$  as
$G^+ \sim \Delta z_0 + 1/N$ and $G^- \sim \Delta z_0 - 1/N$, and thus $(G^+ - G^-)/G_0 \sim
(1/N)/\Delta z_0 \sim 1/\sqrt{N^2 P}$.

As shown in Fig.~\ref{fig:sigma_and_gamma}, the standard deviation of $P(G_1/G_0)$ vanishes for large $N^2 P$ as  $(N^2 P)^{-\beta}$ with $\beta \approx 0.35$, i.e. somewhat slower than predicted. As we will argue now, as long as $\beta > 1/4$, $G$ is still well defined in the thermodynamic limit.

Let us ask the following: Can we estimate the deviation in $G$ in the thermodynamic limit for a fixed strain $\gamma_{t}$? For large $N^2 P$, making and breaking events are equally likely, and as $\gcc \sim \sqrt{P}/N$, the number of these events for fixed strain $\gamma_{t}$ can be estimated to diverge as $N/\sqrt{P}$. Under the assumption that each of these events are drawn independently from a distribution
with a variance that scales as  $(N^2 P)^{-2\beta}$, we find that the variance in $G_1$ is of order $N^{1-4\beta} P^{-1/2-2 \beta}$, which converges to zero in the large $N$ limit when $\beta> 1/4$, as is clearly the case here.
We believe this to be consistent with a picture where, for large systems,
the effective value of
$G$ depends on the strain only, and not on the total number of contact changes \cite{dagoisbohy2014oscillatoryrheologynear, boschaninprep}.

\textit{Discussion:} We now compare our work to recent studies of contact changes
in nonlinearly vibrated
jammed packings \cite{schreck2011repulsivecontactinteractions}.
Consistent with our work, contact changes were found to occur for vanishingly small perturbations
when either $N\rightarrow \infty$, or $P\rightarrow 0$.
Nevertheless, the obtained scaling relations are different.
We note that the procedure used in \cite{schreck2011repulsivecontactinteractions}
is very different: Schreck {\em et al} vibrate their packings and determine the critical
perturbation amplitude by {\em averaging} over all eigenmodes, whereas our protocol
employs a single mode of deformation. Clearly, the conceptually simpler shear deformation
used here will predominantly excite lower frequency modes, and does away with the need
to perform such averages. Perhaps not coincidentally, the experimentally relevant protocol of shear
leads to a much cleaner and clearer scaling result.

We point out several important questions for future work. First, can the
first contact change be predicted
from combining the statistics of overlap (force distribution), underlap (pair correlation function)
and non-affine deformations \cite{ellenbroek2006criticalscalingin, ellenbroek2009jammedfrictionlessdisks,wyart2012marginalstabilityconstrains,saitoh2013masterequationprobability}?
Our preliminary explorations suggest that this may not be the case: for example the first contact break
appears to correspond to an atypical combination of deformation and overlap.
Second, we have started to explore contact changes beyond the first, and have found strong correlations
between subsequent contact changes, which appear to organize in series of break-make events, for which
at present we do not have a clear explanation. 

We finally stress here that even though the first contact change
signals the end of \emph{strict} linear response, its predictions for
macroscopic observables such as the shear modulus remain
relevant far beyond the first contact change. 
A wider implication of our work
is to uncover the unique character of
rearrangements in marginal materials: Microscopic rearrangements in systems in the vicinity of a jamming transition are restricted to the particle scale, which is qualitatively distinct from  denser amorphous systems, which are dominated by collective, avalanche-like events.


We thank L Gómez, S Henkes, CP Goodrich, AJ Liu, SR Nagel and CS O'Hern
for discussions. SDB acknowledges funding
from the Dutch physics foundation FOM, and
MvD, BPT and MvH acknowledge funding from the Netherlands Organization for Scientific Research (NWO).


%

\end{document}